\documentclass[11pt,superscriptaddress,aps,prd,preprint]{revtex4}
\usepackage{epsfig}
\usepackage{verbatim}
\usepackage{amsmath}
\usepackage{graphicx}

\newcommand{\incps}[5]{\includegraphics[#2,#3][#4,#5]{#1}}

\begin{document}
\title{The Slavnov-Taylor Identities for the 2+1 Dimensional Noncommutative CP$^{N-1}$ Model}

\author{B. Charneski}
\affiliation{Instituto de F\'{\i}sica, Universidade de S\~ao Paulo\\
Caixa Postal 66318, 05315-970, S\~ao Paulo, SP, Brazil}
\email{bruno,mgomes,ajsilva@fma.if.usp.br}

\author{M. Gomes}
\affiliation{Instituto de F\'{\i}sica, Universidade de S\~ao Paulo\\
Caixa Postal 66318, 05315-970, S\~ao Paulo, SP, Brazil}
\email{bruno,mgomes,ajsilva@fma.if.usp.br}

\author{T. Mariz}
\affiliation{Instituto de F\'\i sica, Universidade Federal de Alagoas, 57072-270, Macei\'o, Alagoas, Brazil}
\email{tmariz@if.ufal.br}

\author{J. R. Nascimento}
\affiliation{Departamento de F\'{\i}sica, Universidade Federal da Para\'{\i}ba\\
Caixa Postal 5008, 58051-970, Jo\~ao Pessoa, Para\'{\i}ba, Brazil}
\email{jroberto@fisica.ufpb.br}

\author{A. J. da Silva}
\affiliation{Instituto de F\'{\i}sica, Universidade de S\~ao Paulo\\
Caixa Postal 66318, 05315-970, S\~ao Paulo, SP, Brazil}
\email{bruno,mgomes,ajsilva@fma.if.usp.br}

\date{\today}
\begin{abstract}
In the context of the $ 1/N $ expansion, the validity of the Slavnov-Taylor identity relating three and two point functions for the $2+1$ dimensional noncommutative CP$^{N-1} $ model is investigated, up to subleading $1/N$ order, in the Landau gauge. 
\end{abstract}
\maketitle

\section{Introduction}

Historically, the Slavnov-Taylor  (ST) identities \cite{SlavnovTaylor} have played an essential role in proving the renormalizability of non-abelian gauge theories \cite{tHooftVeltman}. It is therefore important
to know the limitations or even the validity of these identities whenever new structures as algebra deformations and space noncommutativity are introduced.  Nowadays, this issue has aroused a deal of attention particularly due to results that seem to indicate that at the Planck's scale the space may become noncommutative \cite{DoplicherFredenhagenRoberts}. In this situation the coordinates should satisfy
\begin{equation}
[x^{\mu},x^{\nu}]=i\theta^{\mu\nu},
\end{equation}
where for the most studied case, called canonical noncommutativity,  $\theta^{\mu\nu}$ is a constant, antisymmetric matrix. In general terms, the unleashing of noncommutativity signals not only for the breaking of Lorentz invariance but also leads to the appearance of an ultraviolet metamorphosis, the so called IR/UV mixing, which may destroy the perturbative scheme \cite{MinwallaRaamsdonkSeiberg}. Besides these basic aspects the possible modifications of results linked to standard symmetries must also be investigated. It has been proved, for example, that CPT symmetry is preserved by the noncommutativity, in spite of its strong nonlocality \cite{SheikhJabbariChaichianNishijimaTureanu}. Gauge symmetry seems also to be
 important to secure the presence of Goldstone bosons for spontaneously broken symmetries \cite{Campbell}. Concerning the  ST identities, exploratory studies have been dedicated to the effects of the noncommutativity on the renormalization of the QED$_4$ \cite{Hayakawa} and also specific scattering processes in the tree approximation \cite{MarizPiresRibeiro}. These studies were complemented by a systematic analysis at the one loop level for QED$_4$ in Ref.~\cite{Charneski}. Such studies are  relevant particularly taking into account the incoming LHC experiments to test possible extensions of the standard model. Going further with these investigations,
  in this work  we shall analyze the possible modifications on the ST identities due to the noncommutativity of the underlying  space in the context  of the three dimensional CP$^{N-1}$ model. When compared with QED$_4$, the new feature in this model is the absence of a kinetic  term for the gauge field, which however is generated by quantum corrections. This study is also a natural sequel of an earlier work on the noncommutative CP$^{N-1}$ model in which, up to the leading order of $1/N$,  the absence of dangerous UV/IR mixing was proved \cite{AsanoGomesRodriguesSilva}.
 
The noncommutative CP$^{N-1}$ model is defined by the action
\begin{eqnarray}\label{eq3}
S &=& \int{d^3x}\Bigg{\{}\partial^{\mu}\phi_{a}^{\dagger}\partial_{\mu}\phi_{a}
-m^2\phi_a^{\dagger}\star \phi_a+\lambda\star \Big(\phi_a\star\phi_a^{\dagger}-{\frac{N}{g}}\Big)
\nonumber\\
&+&e^2\phi_a^{\dagger}\star A^{\mu}\star A_{\mu}\star \phi_a+ie\Big(\partial^{\mu}\phi_a^{\dagger}\star A_{\mu}\star \phi_a-\phi_a^{\dagger}\star A_{\mu}\star \partial_{\mu}\phi_a\Big)
\\
&-&{\frac{N}{2\alpha}}\,(\partial^{\mu}A_{\mu})\star (\partial^{\nu}A_{\nu})+N\partial^{\mu}\bar{c}\star \Big[\partial_{\mu}c-ie\big(c\star A_{\mu}-A_{\mu}\star c\big)\Big]\Bigg{\}}~,\nonumber
\end{eqnarray}
where $\phi_a$  $(a=1,...,N)$ is a $N$-tuple of charged scalar fields transforming in accord with the left fundamental representation of the $U_{\star }(1)$ group,
\begin{equation}
\phi_{a}(x)\quad\rightarrow\quad U_{\star }(x)\star  \phi_{a}(x), \qquad U_{\star }(x)={\rm e}^{i\Lambda(x)}_{\star }\equiv 1 +i\Lambda(x)-\frac12 \Lambda(x)\star  \Lambda(x)+\cdots,
\end{equation}
the star symbol denoting the Moyal product (for a review about noncommutativity see \cite{NekrasovDouglasSzabo})
\begin{eqnarray}\label{eq24}
f(x)\star g(x)=e^{(i/2)\Theta^{\mu\nu}\partial_{_{x}\mu}\partial_{_{y}\nu}}f(x)g(y)\Big{|}_{x=y}~.
\end{eqnarray}

Besides the gauge field, the auxiliary field $ \lambda $, which implements the constraint $\phi_a\star \phi_a^{\dagger}={\frac{N}{g}}$, is taken in the adjoint representation of the gauge group, i.e.,
\begin{equation}
\lambda(x)\quad\rightarrow\quad\lambda^\prime(x) = U_{\star }(x)\star\lambda(x)\star U_{\star}^{-1}(x).
\end{equation}
The great advantage of this choice is that $ \lambda $ and $ A_{\mu} $  are then, in the leading $ 1/N $ order, independent fields.  In the present situation, the propagators will be given by 
\begin{subequations}
\begin{eqnarray}
\label{propa}\raisebox{-0.4cm}{\incps{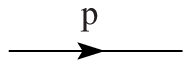}{-1.5cm}{-.5cm}{1.5cm}{.5cm}}
 &=& \Delta^0(p)=\frac{i}{p^2-m^2}, \\
\label{propb}\raisebox{-0.0cm}{\includegraphics{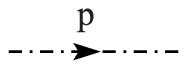}}
\!\!\!\qquad &=& S^0(p)=\frac{i}{Np^2},\\
\label{propc}\raisebox{-0.4cm}{\incps{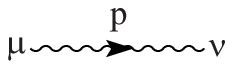}{-1.5cm}{-.5cm}{1.5cm}{.5cm}} &\approx & D^0_{\mu\nu}=  \frac{16i}{Ne^2}\left(g_{\mu \nu }-\frac{p_{\mu }p_{\nu }}{p^{2}}\right)\left(\frac{1}{\sqrt{-p^{2}}}-\frac{4m}{\pi p^2}\right)-\frac{i\alpha}{N}\frac{p_{\mu }p_{\nu }}{p^{4}}, \label{propagadorb}\\
\label{propd}\raisebox{-0.0cm}{\includegraphics{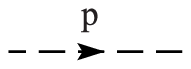}}
\!\!\!\qquad  &\approx & \Delta^0_\lambda(p) = \frac{8i\sqrt{-p^{2}}}{N}\left(1+\frac{4m}{\pi}\frac{1}
{\sqrt{-p^ 2}}\right).
\end{eqnarray}
\end{subequations}
Note that the propagators (\ref{propa}) and (\ref{propb}) are obtained directly from the action (\ref{eq3}) considering the quadratic part of the fields $\phi$ and $c$, whereas  the propagators for the gauge (\ref{propc}) and auxiliary (\ref{propd}) fields are obtained perturbatively, by considering large spacelike $p$ behavior. 

\begin{figure}[h]
\begin{center}\includegraphics{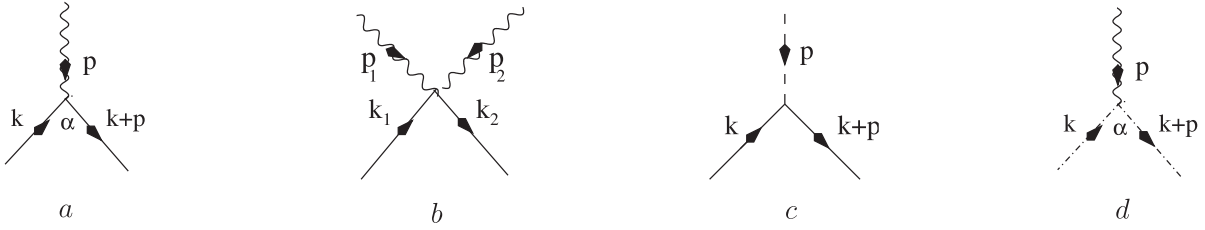}
\end{center}
\caption{Vertices associated to the action (\ref{eq3}).
}
\label{FigVerCPN}
\end{figure}

The vertices for the theory are the following
\begin{subequations}
\begin{eqnarray}
ie\Big(\partial^{\mu}\phi_a^{\dagger}\star A_{\mu}\star \phi_a-\phi_a^{\dagger}\star A_{\mu}\star \partial_{\mu}\phi_a\Big)\qquad &\leftrightarrow &\qquad -ie(2k+p)_{\mu}{\textrm{e}}^{-ik\wedge p}\label{n7}\\
e^2\phi_a^{\dagger}\star A^{\mu}\star A_{\mu}\star \phi_a\qquad &\leftrightarrow &\qquad 2ie^2g^{\mu\nu}{\textrm{e}}^{-ik_{1}\wedge k_{2}}
\cos (p_{1}\wedge p_{2})\label{n8}\\
 \lambda \star\phi_a\star\phi_a^{\dag}\qquad &\leftrightarrow &\qquad i{\textrm{e}}^{-ik\wedge p}
\label{n9}\\
-ieN\partial^\mu\bar{c}\star\left(c\star A_{\mu}-A_{\mu}\star c\right)\qquad &\leftrightarrow &\qquad 2eNk^{\alpha}\sin(p\wedge k)\label{n9a},
\end{eqnarray}
\end{subequations}
such that the graphical representation are given respectively in Fig.~(\ref{FigVerCPN}).
 
Notice that, as indicated in the last line of (\ref{eq3}), we are adopting a generic Lorentz gauge fixing whereas the calculations performed in
 \cite{AsanoGomesRodriguesSilva} were restricted to the Landau gauge. Our gauge fixing together  with the term for the ghost fields $c$ and $ \bar c $ signalize a formal symmetry associated with the invariance of the action under BRST transformations which have the following form 
\begin{subequations}\label{eq4}
\begin{eqnarray}
\phi_a &\rightarrow& \phi_a^{\prime}=\phi_a+ic\star \phi_a\,\epsilon~,\\
\phi^{\dagger}_a &\rightarrow& \phi_a^{\dagger\prime}=\phi_a^{\dagger}-i\phi_a^{\dagger}\star \,c\,\epsilon~,\\[5pt]
A_{\mu} &\rightarrow& A_{\mu}^{\prime}=A_{\mu}-\frac{1}{e}\,\partial_{\mu}c\,\epsilon+i[c,A_{\mu}]_{\star }\,\epsilon~,\\[5pt]
\lambda &\rightarrow& \lambda'=\lambda+i[c,\lambda]_{\star }\,\epsilon~,\\
c &\rightarrow& c^{\prime}=c-ic\star c\,\epsilon~,\\[5pt]
\bar{c} &\rightarrow& \bar{c}^{\prime}=\bar{c}-\frac{1}{e\alpha}\,\partial^{\mu}A_{\mu}\,\epsilon~,
\end{eqnarray}
\end{subequations}
where $\epsilon$ is an infinitesimal Grassmannian  parameter. Due to the presence of the Moyal product, the implications of this invariance have to be examined anew. In particular we shall inspect  the ST identities characteristics of this invariance but, as the leading contributions in $ 1/N $ involve both the one-loop and two-loop diagrams whose analytic expressions are very intricate, we will focus
directly on the asymptotic behavior for high momenta of the relevant Green functions.

To derive the ST identities, as usual, we add to the source terms for the basic fields the source terms associated to the BRST transformations
\begin{eqnarray}\label{eq6}
&S&\!\!\!_{source}=\int{d^3x}\Bigg{\{} J_{\mu}\star A^{\mu}+\eta^{\dagger}_a\star \phi_a+\phi^{\dagger}_a\star \eta_a+\bar\xi \star c+\bar{c}\star \xi
+\zeta \star \lambda+u\star (i\,[c,\lambda]_{\star })\nonumber\\
&+&\!\!K_{\mu}\star \Big(-\frac{1}{e}\,\partial^{\mu}c+i[c,A^{\mu}]_{\star }\Big)+v\star (-ic\star c)+\omega^{\dagger}_a\star (ic\star \phi_a)
+(-i\phi^{\dagger}_a\star c)\star \omega_a\Bigg{\}}.
\end{eqnarray}
\noindent
The invariance of the functional generator under the field transformations (\ref{eq4}) formally allows for the  ST identity
\begin{eqnarray}\label{eq7}
\int{d^3x}\Bigg{\{}J_{\mu}\star {\delta W\over\delta
K_{\mu}}+\eta^{\dagger}_{a}\star {\delta
W\over\delta\omega^{\dagger}_a} -{\delta
W\over\delta\omega_a}\star \eta_a+\bar\xi \star {\delta W\over\delta v}
+\frac{1}{e\alpha}\,\partial_{\mu}{\delta W\over\delta
J_{\mu}}\star \xi+\zeta\star {\delta
W\over\delta u}\Bigg{\}}=0~,
\end{eqnarray}
where $W$ is the  functional generator for the connected Green functions. The above result together with the relation
\begin{eqnarray}\label{eq8}
\int{d^3x}\left(i\xi + Ne\,\partial_{\mu}{\delta W\over\delta K_{\mu}}\right)=0,
\end{eqnarray}

\noindent obtained from the invariance of $S+S_{source}$ under a general transformation $\delta\bar{c}$ of the ghost field, constitutes a powerful tool for the study of the UV behavior of field theories. 

We begin the analysis of the above identities by proving that the longitudinal part of the gauge field propagator is not modified by radiative corrections, as it happens in \cite{Charneski}. In fact, by functionally deriving  (\ref{eq7}) with respect to  the $J^{\nu}(y)$ and $\xi(z)$ sources, we get
\begin{eqnarray}\label{eq14}
{\delta^2 W\over\delta\xi(z)\delta K^{\nu}(y)}\left \vert\,+\,\frac{1}{e\alpha}\,\partial_{z}^{\mu}{\delta^2 W\over\delta J^{\nu}(y)\delta J^{\mu}(z)}\right \vert\,=0,
\end{eqnarray}
where henceforth a vertical bar is used just to remember that the function immediately to its left must be calculated with all sources equal to zero.
Now, from  (\ref{eq8}) it follows that
\begin{eqnarray}\label{eq16}
\left. Ne\,\partial_{x}^{\mu}{\delta^2 W\over\delta\xi(z)\delta K^{\mu}(x)}\right \vert=-i\delta^3(x-z)~,
\end{eqnarray}
implying that
\begin{eqnarray}\label{eq18}
-\left. \frac{i}{N}\,\delta^3(y-z)+{\frac{1}{\alpha}\,}\partial_{y}^{\nu}\partial_{z}^{\mu}{\delta^2 W\over\delta J^{\nu}(y)\delta J^{\mu}(z)}\right \vert=0~.
\end{eqnarray}
In momentum space, this equation becomes: 
\begin{eqnarray}\label{eq21}
k^{\mu}k^{\nu}D_{\mu\nu}(k)=-\frac{i\alpha}{N}~,
\end{eqnarray}
so that the longitudinal part of the gauge propagator, which is proportional to $k_\mu k_\nu$, must be given by
\begin{eqnarray}
D^{L}_{\mu\nu}(k)=-\frac{i\alpha}{N}\,\frac{k^{\mu}k^{\nu}}{(k^2)^2}~.
\label{parlongpropfoton}
\end{eqnarray}
Therefore, at any finite order of $ 1/N $, it  is not affected by the noncommutativity. This result will be used in the forthcoming analysis of the ST identity.

We now consider the three point function which involves the gauge and the charged fields. $\langle 0|T A^{\mu}\phi\phi^{\dagger}|0\rangle$ by deriving  (\ref{eq7})  with respect to the sources $\eta_a(x)$,
$\eta^{\dagger}_b(y)$ and $\xi(z)$, we get
\begin{eqnarray}\label{eq22}\left.
\frac{\delta^3 W}
{\delta\xi (z)\delta\eta_a (x)\delta\omega^{\dagger}_{b}(y)}\right\vert -
\left.\frac{\delta^3
W}{\delta\xi(z)\delta\eta^{\dagger}_b(y)\delta\omega_a(x)}\right \vert +
\frac{1}{e\alpha}\,\partial_{z}^{\mu}\left.\frac{\delta^3
W}{\delta\eta^{\dagger}_b(y)\delta\eta_a(x)\delta
J^{\mu}(z)}\right\vert=0
\end{eqnarray}
or, equivalently,
\begin{eqnarray}\label{eq23}
\!\!\frac{1}{e\alpha}\,\partial_{z}^{\mu}\langle{\rm T}\phi_b(y)\phi^{\dagger}_a(x)A_{\mu}(z)\rangle=
i\langle {\rm T}\bar{c}(z)\phi^{\dagger}_a(x)\,c(y)\star \phi_b(y)\rangle
-i\langle {\rm T}\bar{c}(z)\phi_b(y)\,\phi^{\dagger}_a(x)\star c(x)\rangle.
\end{eqnarray}

It is convenient to write the above identity in terms of the one-particle irreducible vertex functions whose generating functional, $ \varGamma $, is defined by
\begin{eqnarray}
W[J,\eta,\bar\eta,\xi,\bar\xi;K,v,\omega,\bar\omega] &&= \varGamma[A_{cl},\phi_{cl},\phi^{\dagger}_{cl},C_{cl},\bar C_{cl};K,v,\omega,\bar\omega] \nonumber\\
&&+ \int d^4x \left( J_\mu \star  A^{\mu}_{cl} + \eta^{\dagger} \star  \phi_{cl} + \phi^{\dagger}_{cl} \star  \eta + \bar\xi \star  C_{cl} + \bar C_{cl} \star  \xi \right),
\end{eqnarray}
where we have introduced the classical fields
\begin{equation}
 A^{\mu}_{cl}= \frac{\delta W}{\delta J_\mu}~,\qquad  \phi_{cl} = \frac{\delta W}{\delta  \eta^{\dagger}}~,\qquad  \phi^{\dagger}_{cl} =- \frac{\delta W}{\delta  \eta}~,\qquad C_{cl}=\frac{\delta W}{\delta \bar \xi}~, \qquad \bar C_{cl}=-\frac{\delta W}{\delta  \xi}.
\end{equation}

Employing the momenta representation, it then follows that
\begin{eqnarray}\label{eq27a}
\frac{i}{e\alpha}&&\!\!\!\!\!\!\!\!\!(p_{3})_{\mu}D^{\mu\nu}(p_3)\Delta(p_2)\Delta(p_1)\Gamma_{\nu}(p_2,-p_1,p_3)\nonumber\\
&=&i\int\frac{d^3k}{(2\pi)^3}~e^{ik\wedge p_2}~\Delta(p_1)\Delta(k)S(p_2-k)S(-p_3)\Gamma_4(k,-p_1,p_2-k,p_3)\label{eq28}\\
&-&i\int\frac{d^3k}{(2\pi)^3}~e^{-ik\wedge p_1}~\Delta(p_2)\Delta(k)S(-p_1+k)S(-p_3)\Gamma_4(p_2,-k,-p_1+k,p_3),\nonumber
\end{eqnarray}
where in a simplified notation $S(k)$ and $\Delta(k)$ represent the Fourier transforms of $S(x)$ and $\Delta(x)$, respectively the matter field and the ghost field  propagators. The $\Gamma$ functions introduced above are the Fourier transforms of
\begin{eqnarray}\label{eq27}
\Gamma_{\nu}(a,x;b,y;z)&=&
{\delta^3\varGamma\over\delta\phi^{\dagger}_a(x)\delta\phi_b(y)\delta A^{\nu}_{cl}(z)},\\
\Gamma_4(a,x;b,y;z;u)&=&
{\delta^4\varGamma\over\delta\phi^{\dagger}_a(x)\delta\phi_b(y)\delta\bar{c}_{cl}(z)\delta{c}_{cl}(u)}.
\end{eqnarray}

The steps leading to (\ref{eq27a}) are very formal but its validity can be directly verified as we shall do now, up to the subleading order of $1/N$. We note that this equation can be rewritten as
\begin{eqnarray}\label{eq29}
\frac{1}{Ne}\frac{p^{\nu}_{3}}{p_3^2}\Gamma_{\nu}(p_2,-p_1,p_3)
=\Delta^{-1}(p_2)S(-p_3)H_2(p_1,p_2,p_3)-\Delta^{-1}(p_1)S(-p_3)H_1(p_1,p_2,p_3),
\end{eqnarray}
where we have used the identity (\ref{parlongpropfoton}) for the longitudinal part of the gauge field propagator and, as suggested in an analysis of the ST identities for QCD \cite{MarcianoPagels}, introduced the functions 
\begin{eqnarray}
H_1(p_1,p_2,p_3)&=&i\int\frac{d^3k}{(2\pi)^3}~e^{-ik\wedge p_1}~\Delta(k)S(-p_1+k)\Gamma_4(p_2,-k,-p_1+k,p_3),
\label{defH1}\\[5mm]
H_2(p_1,p_2,p_3)&=&i\int\frac{d^3k}{(2\pi)^3}~e^{ik\wedge p_2}~\Delta(k)S(p_2-k)\Gamma_4(k,-p_1,p_2-k,p_3).
\label{defH2}
\end{eqnarray}

We will now check (\ref{eq29}) up to subleading order of $ 1/N $. Note first that, including corrections up to $ 1/N $ order, the matter field propagator is given by
\begin{eqnarray}
\Delta(p)=\frac{i}{p^2-m^2-\frac{i}{N}\Sigma_{\phi}(p)}.
\label{propesccorriult}
\end{eqnarray}
From now on, we will work in the Landau gauge, $\alpha=0$. Adopting dimensional regularization with minimal subtraction, we have
\begin{eqnarray}\label{autoenergiacampoescalar}
\Sigma_{\phi}^{unr}(p)&=& -N\,\int\frac{d^Dk}{(2\pi)^D}\,(k+2p)^{\mu}D^0_{\mu\nu}(k)(k+2p)^{\nu}\Delta^0(k+p)-N\int\frac{d^Dk}{(2\pi)^D}\,\Delta^0(k+p)\,\Delta^0_{\lambda}(k)\nonumber\\
&=&-\frac{20i}{\pi^2}\,\frac{1}{\epsilon}\,p^2 + \mathrm{finite\;terms},
\end{eqnarray}
where the superscripts $unr$ denotes unrenormalized function. The convenient counterterm is $b\,\partial_{\mu}\phi_a^{\dagger}\partial^{\mu}\phi_a$, where the renormalization constant is $b=\frac{20}{N\pi^2}\,\frac{1}{\epsilon}$. As for the ghost propagator, we obtain
\begin{eqnarray}
S(p_3)=\frac{i}{p_3^2\left[N-i\Sigma_{c}(p_3)\right]}.
\label{propfantult}
\end{eqnarray}
The unrenormalized $\Sigma_{c}(p_3)$ is given by
\begin{eqnarray}\label{correcpropfant}
\Sigma_{c}^{unr}(p_3)&=&\left(\frac{1}{p_3^2}\right)\left[-(2eN)^2\int\frac{d^Dk}{(2\pi)^D}\,(k+p_3)^{\mu}D^0_{\mu\nu}(k)p_3^{\nu}\,S^0(k+p_3)\sin^2(k\wedge p_3)\right].
\label{correcpropfant}
\end{eqnarray}
The result for the planar part is
\begin{eqnarray}
\Sigma_{c}^{unr}(p_3)= -\frac{32i}{3\pi^2}\,\frac{1}{\epsilon}+\mathrm{finite\;terms},
\end{eqnarray}
which may be renormalized by the counterterm $fN\partial_{\mu}\bar{c}\,\partial^{\mu}c$, with $f=\frac{32}{3N\pi^2}\frac1{\epsilon}$.

The unrenormalized three point vertex $\Gamma_\nu$ and $H_{m}$ functions have the following expansions
\begin{eqnarray}
\Gamma_{\nu}=\Gamma_{\nu}^{0}+\frac1N\Gamma_{\nu}^{1unr}
\label{funcvertcorrig}
\end{eqnarray}
and
\begin{eqnarray}
H_{m} = H^{0}_{m} +  \frac1N H^{1}_{m},
\label{funcHcorrig}
\end{eqnarray} 
up to $1/N$ order. We have verified that the $H^1_{m}$ functions are not UV divergent, therefore no counterterms are needed. However, as shown in \cite{AsanoGomesRodriguesSilva}, $\Gamma_{\nu}^{1unr}$ consists of divergent diagrams with one and two loops. In the two loop case, the regularization is introduced just in the last integral. Thus,  the total UV divergence is given by
\begin{eqnarray}\label{divfuncvert1N}
\Gamma_{\nu}^{1unr}=\frac{28ie(2p_2+p_3)_{\nu}}{3\pi^2}\,\frac{1}{\epsilon}.
\end{eqnarray}
The numerical difference, a factor of two, from Ref.~\cite{AsanoGomesRodriguesSilva}, is due to a different regularization prescription adopted in that work. Therefore, the counterterm term is
\begin{eqnarray}
B\,ie\Big(\partial^{\mu}\phi_a^{\dagger}\star A_{\mu}\star \phi_a-\phi_a^{\dagger}\star A_{\mu}\star \partial_{\mu}\phi_a\Big),
\end{eqnarray}
where the renormalization constant is $B=\frac{28}{3N\pi^2}\frac{1}{\epsilon}$. 

Using the above notation, and allowing terms up to $1/N$ order, the identity (\ref{eq29}) may be rewritten as
\begin{eqnarray}
\frac{1}{Ne}\left(\frac{p_3^{\nu}}{p_3^2}\right)\left[\Gamma_{\nu}^{0}+\frac{1}{N}\Gamma_{\nu}^{1}\right]\left[p_3^2\left(N-i\Sigma_{c}(p_3)\right)\right]\nonumber\\[3mm]=\left\{\left[p_2^2-m^2-\frac{i}{N}\Sigma_{\phi}(p_2)\right] \left[H^{0}_{2} +  \frac1N H^{1}_{2}\right]\right\}
\nonumber\\[3mm]
-\left\{\left[p_1^2-m^2-\frac{i}{N}\Sigma_{\phi}(p_1)\right] \left[H^{0}_{1} +  \frac1N H^{1}_{1}\right]\right\},
\label{identicomtermosexpan}
\end{eqnarray}
where the renormalized functions are given by
\begin{subequations}\label{fr}
\begin{eqnarray}
\Gamma_{\nu}^{1} &=& \Gamma_{\nu}^{1unr} + NB\,\Gamma_{\nu}^{0}, \\
\Sigma_{c} &=& \Sigma_{c}^{unr} + iNf, \\
\Sigma_{\phi} &=& \Sigma_{\phi}^{unr} + iNb.
\end{eqnarray}
\end{subequations}


To obtain the ST identity at leading order we must consider the vertex function $\Gamma^{0}_{\nu}$ on the left hand side of the expression (\ref{identicomtermosexpan}). The right hand side receives the contribution of $H^{0}_{1}$ and $H^{0}_{2}$, which are both equal to $ie^{-ip_2\wedge p_3}$. Replacing these results in (\ref{identicomtermosexpan}), we get
\begin{eqnarray}
\frac{1}{e}\,p_3^{\nu}(-ie)(2p_2+p_3)_{\nu}\,e^{-ip_2\wedge p_3}=(p_2^2-m^2)ie^{-ip_2\wedge p_3}-(p_1^2-m^2)ie^{-ip_2\wedge p_3},
\end{eqnarray}
which is identically satisfied, as can be seen by using the momentum conservation $p_1=p_2+p_3$. 


Less trivial result is obtained when we analyze the subleading order which receives loop corrections. As we will see, the identity in subleading order carries quantum corrections and establish a relation among the renormalization constants. Therefore, from (\ref{identicomtermosexpan}) we must have
\begin{eqnarray}
\frac{1}{e}\,p_3^{\nu}\left[\Gamma_{\nu}^{1}-i\,\Gamma_{\nu}^{0}\Sigma_{c}(p_3)\right]&=&\left[\left(p_2^2-m^2\right)H^1_2-iH^0_2\Sigma_{\phi}(p_2)\right]\nonumber\\
&-&\left[\left(p_1^2-m^2\right)H^1_1-iH^0_1\Sigma_{\phi}(p_1)\right].
\label{ordem1expans}
\end{eqnarray}
Replacing (\ref{fr}) into the above expression, the UV divergences of the unrenormalized functions shown in (\ref{fr}) cancel each other, which proves the validity of the noncommutative ST identities for the CP$^{N-1}$ model. Furthermore, we obtain the relation involving the renormalization constants, $B+f=b$.

\section{Conclusion}\label{Conclusao}

We have verified the ST identity in the $1/N$ expansion for the noncommutative CP$^{N-1}$ model. As is known, the diagrams of $1/N$ order involve one and two loops which are very intricate. Therefore, we restricted ourselves to the verification of the matching of the UV divergent parts. Besides these UV parts, we have also infrared (IR) singular parts coming from nonplanar parts of the functions. However, in \cite{AsanoGomesRodriguesSilva} it was shown that the leading IR singular parts are canceled due to diagramatic identities, leaving only logarithmic singularities, which are not problematic as they are integrable.

{\bf Acknowledgements.} This work was partially supported by Conselho
Nacional de Desenvolvimento Cient\'{\i}fico e Tecnol\'{o}gico (CNPq),
Funda\c{c}\~{a}o de Amparo \`{a} Pesquisa do Estado de S\~{a}o
Paulo (FAPESP), and Coordena\c{c}\~{a}o de Aperfei\c{c}oamento do Pessoal
do Nivel Superior (CAPES).

\end{document}